\documentclass[a4paper,12pt]{article}

\usepackage{fancyhdr}
\usepackage{amssymb}
\usepackage{tikz}
\usepackage{graphicx}
\usepackage[colorlinks,hyperfootnotes=false]{hyperref}

\newcommand{\fn}[2]{\mathinner{#1\mathopen{\left(#2\right)}}}

\newcommand{\eq}[1]{Eq.~(\ref{#1})}
\newcommand{\eqs}[2]{Eqs.~(\ref{#1}) and (\ref{#2})}
\newcommand{\eqss}[3]{Eqs.~(\ref{#1}), (\ref{#2}) and (\ref{#3})}

\newcommand{\eV}{\mathinner{\mathrm{eV}}}

\newcommand{\MeV}{\mathinner{\mathrm{MeV}}}
\newcommand{\GeV}{\mathinner{\mathrm{GeV}}}
\newcommand{\TeV}{\mathinner{\mathrm{TeV}}}

\newcommand{\axino}{{\tilde{a}}}

\newcommand{\planck}{\mathrm{Pl}}

\def\bea{\begin{eqnarray}}
\def\eea{\end{eqnarray}}
\def\beq{\begin{equation}}
\def\eeq{\end{equation}}

\begin{document}

\title{A simple model for particle physics and cosmology}
\author{Wan-Il Park \\[2ex] \textit{Department of Physics, KAIST, Daejeon 305-701, South Korea} \\}

\maketitle

\abstract{
We propose a simple extension of the minimal supersymmetric standard model by introducing a gauge singlet in addition to right-handed neutrinos.
The model resolves the strong CP problem by Pecci-Quinn symmetry, explains the origin of left-handed neutrino masses as well as MSSM $\mu$-parameter.
It also gives rise to thermal inflation, baryogenesis and dark matter in a remarkably consistent way.
Interestingly, resolution of moduli problem by thermal inflation constrains tightly axion coupling constant and flaton decay temperature to be $f_a \sim 10^{12} \GeV$ and $T_\mathrm{d} \sim 100 \MeV$, respectively.
Model parameters in this case are likely to give right amount of baryon asymmetry and dark matter at present.
The main component of dark matter is expected to be the axino whose mass is nearly fixed to be about $1 \GeV$.
}

\thispagestyle{fancy}
\rhead{KAIST-TH 2010/06}

\section{Introduction}
\label{intro}

Although its great success in describing particle physics, the standard model (SM) of particle physics has been faced on some big questions: hierarchy problem associated with the unnatural stability of weak scale Higgs mass against large quantum corrections induced by the large hierarchy between electroweak ($m_\mathrm{ew} \sim 10^3 \GeV$) and Planck ($M_\planck  \sim 10^{19} \GeV$) scales, strong CP problem \cite{Kim:1986ax} and the origin of left-handed neutrino masses which is implied by the neutrino oscillation \cite{Bilenky:1978nj}.
Any plausible theory for particle physics, which is based on SM, should address all these questions.

The minimal supersymmetric standard model (MSSM) is a very natural and simple supersymmetric extension of standard model, in which large hierarchy problem is absent by the help of supersymmetry (SUSY) \cite{Nilles:1983ge}, hence it is quite attractive.
MSSM may be the right direction to a theory for low energy physics, but it has a mysterious parameter $\mu$, the dimensionful coefficient of Higgs bilinear superpotential term, whose origin need to be explained \cite{Kim:1983dt}.
In addition, MSSM still should be extended to resolve strong CP problem and explain the origin of left-handed neutrino masses.

Meanwhile, string/M theory, which may be the fundamental theory at high energy, predicts the existence of light particles with gravitationally suppressed interactions: gravitinos and moduli/modulinos.
Typically, those particles are expected to be produced too much due to large reheating temperature or coherent oscillation after primordial inflation, disturbing the successful Big-Bang nucleosynthesis (BBN) or over-closing the present universe \cite{Khlopov:1984pf,Coughlan:1983ci}.
The most compelling solution to this gravitino/moduli problem is thermal inflation \cite{Lyth:1995hj} which is very likely to occur in the framework of SUSY.
But thermal inflation invalidates most of known baryogenesis mechanisms, which work typically above or at the electroweak scale, by diluting out pre-existing baryon/lepton asymmetry 
\footnote{
One may think that Affleck-Dine baryogenesis in a scenario of gauge-mediated SUSY-breaking can work well even in the presence of thermal inflation \cite{de Gouvea:1997tn}.
However, the formation of Q-balls makes it difficult to work \cite{Kasuya:2001tp}.
}.
Since the reheating temperature after thermal inflation is typically well below the electroweak scale, baryogenesis mechanism is difficult to work in general.

In this paper, we propose a simple extension of MSSM by introducing just a gauge singlet and right-handed neutrino chiral superfields, but without any ad-hoc mass parameter except soft SUSY-breaking ones. 
In the framework of gravity-mediated SUSY-breaking, the model realizes Peccei-Quinn symmetry \cite{Peccei:1977hh,Kim:1979if,Zhitnitsky:1980tq} to resolve strong CP problem and explains the origin of left-handed neutrino masses through see-saw mechanism \cite{seesaw} as well as that of MSSM $\mu$-parameter in a very efficient way.
The model also provides remarkably consistent cosmology including thermal inflation, baryogenesis and dark matter.

This paper is organized as follows.
In Section 2, we propose a model and describe how it provides proper low energy physics.
In Section 3, we briefly describe cosmology which the model can realize.
In Section 4, we conclude.

\section{The model}
Motivated by the drawbacks of MSSM, strong CP problem, the origin of left-handed neutrino masses and $\mu$ parameter, we extend the MSSM to the following, assuming SUSY-breaking is mediated by gravitationally suppressed interaction
\footnote{
In the scenario of gravity-mediated SUSY-breaking, the symmetry breaking scale of thermal inflation required to resolve moduli problem coincides accidentally with that of Peccei-Quinn symmetry.
Hence, in the scope of this paper, it is natural to consider such a mediation scenario as the framework of our argument.
We refer the reader to ref. \cite{Choi:2009qd} for the case of mirage-mediation of SUSY-breaking \cite{Endo:2005uy}, and will discuss the cases of other mediation scenarios, for example gauge-mediation \cite{Dine:1981gu,Dine:1994vc}, in other place.
} 
\footnote{
For the MSSM $\mu$-term,  we can use a renormalizable interaction $\lambda_\mu \Phi H_u H_d$ instead of the non-renormalizable interaction $\lambda_\mu \Phi^2 H_u H_d / M_\planck$ which we are using in this paper. 
Our model then involves only renormalizable interactions though $\lambda_\mu \sim 10^{-9}$ is hierarchically small.
It may be still plausible to consider such a hierarchically small Yukawa coupling, since Yukawa couplings of MSSM already have a hierarchy.
Compared to the case of non-renormalizable interaction for $\mu$-term, the physics is barely affected and can be applied to most of known mediation mechanisms of SUSY-breaking except gauge mediation.
}
:
\beq \label{W}
W 
= \lambda_u Q H_u \bar{u} + \lambda_d Q H_d \bar{d} + \lambda_e L H_d \bar{e} + \lambda_\nu L H_u \nu 
+ \lambda_\mu \frac{\Phi^2 H_u H_d}{M_\planck} + \frac{1}{2} \lambda_\Phi \Phi \nu^2
\eeq
with an assumption
\beq \label{mLHusqcond}
m_L^2 + m_{H_u}^2 < 0
\eeq
where $m_L^2$ and $m_{H_u}^2$ are respectively the soft mass-squared parameters of $L$ and $H_u$ around electroweak scale.  
In \eq{W}, indices for gauge group and family structure has been omitted for simplicity, $M_\planck = 2.4 \times 10^{18} \GeV$ is the reduced Planck mass, $\Phi = \phi + \sqrt{2} \theta \axino + \cdots$ is a gauge singlet chiral superfield  and $\nu$ is the right-handed neutrino superfield.
Various new Yukawa couplings are constrained by low-energy physics which will be addressed shortly. 
Note that the model has an accidental $U(1)$ symmetry.
Normalized such that $\Phi$ has charge one, the charges assigned to fields are as follows.
\beq
U(1)\left\{Q, L, H_u, H_d, \bar{u}, \bar{d}, \bar{e}, \Phi, \nu \right\}=\left\{1, 3/2, -1, -1, 0, 0, -1/2, 1, -1/2 \right\}
\eeq

The key feature of our model is that if $\lambda_\Phi \sim \mathcal{O}(1)$ the associated Yukawa interaction can drive the soft mass-squared parameters of $\phi$ and right-handed sneutrino (denoted as $\nu$ except cases of confusion) to be negative through renormalization group running \footnote{See Appendix for renormalization group equations of relevant parameters.}.
If $\nu$ develops vacuum expectation value (vev) before $\phi$ does, our model becomes inconsistent with low energy phenomenology since MSSM $\mu$-term is not reproduced.
In order to avoid this disaster, we assume at Planck scale
\beq
\begin{array}{ccc}
m_\phi^2 \ll m_\nu^2 + |A_\Phi|^2 &,& m_\nu^2 \gtrsim |A_\Phi|^2
\end{array} 
\eeq
where $m_\phi^2$ and $m_\nu^2$ are respectively the soft mass-squared parameters of $\phi$ and $\nu$, and $A_\Phi$ is the $A$-parameter of the trilinear coupling associated with $\lambda_\Phi$. 
In this case, only $\phi$ can develop non-zero vev and the spontaneously broken accidental $U(1)$ symmetry can be identified as the Peccei-Quinn (PQ) symmetry for DFSZ axion \cite{Zhitnitsky:1980tq}.

The potential along $\phi$ is of the form
\beq \label{Vphi}
V(|\phi|) \simeq V_0 + m_\phi^2(Q=|\lambda_\Phi \phi|) |\phi|^2 
\eeq
where $m_\phi^2(Q)$ is the running soft mass-squared of $\phi$ evaluated at a renormalization scale $Q$.
Now 
\bea
\frac{d V}{d |\phi|} &=& \left[ \left( 2 + \frac{d}{d \ln |\phi|} \right) m_\phi^2(|\phi|) \right] |\phi_0|
\\
\frac{d^2 V}{d |\phi|^2} &=& \left[ \left( 1 + \frac{d}{d \ln |\phi|} \right) \left( 2 + \frac{d}{d \ln |\phi|} \right) m_\phi^2(|\phi|) \right]
\eea
and the renormalization group equation of $m_\phi^2$ is 
\beq \label{phimasssqRGE}
\frac{d m_\phi^2}{d \ln Q} = \frac{1}{8 \pi^2} |\lambda_\Phi|^2 \left( m_\phi^2 + m_\nu^2 + |A_\Phi|^2 \right)
\eeq
hence at vacuum where $d V / d |\phi| = 0$ we find 
\beq
m_\phi(\phi_0) = - \frac{1}{16 \pi^2} |\lambda_\Phi|^2 \left( m_\nu^2 + |A_\Phi|^2 \right)
\eeq
where the parameters at right-hand side are evaluated at $Q = |\lambda_\Phi \phi_0|$ and $\phi_0$ is the vacuum position of $\phi$.
Since the renormalization group runnings of $\lambda_\Phi$, $m_\nu^2$ and $|A_\Phi|^2$ are rather slow as shown in Appendix, from \eq{phimasssqRGE} we may approximate $m_\phi^2(\phi_0)$ crudely as 
\beq
m_\phi^2(\phi_0) \sim m_\phi^2(M_\planck) + \frac{1}{8 \pi^2} |\lambda_\Phi|^2 \left( m_\nu^2 + |A_\Phi|^2 \right) \ln \frac{|\lambda_\Phi \phi_0|}{M_\planck}
\eeq
Then the vacuum value $\phi_0$, which is assumed to be real without loss of generality, is
\beq \label{phi0}
\phi_0 \sim \frac{M_\planck}{\lambda_\Phi} \exp \left[ -\frac{1}{2} - \frac{1}{\alpha} \right]
\eeq
where
\beq \label{alpha}
\alpha \equiv \frac{|\lambda_\Phi|^2}{8 \pi^2} \frac{m_\nu^2 + |A_\Phi|^2}{m_\phi^2(M_\planck)}
\eeq
and requiring zero cosmological constant, one finds
\beq \label{V0}
V_0 = \frac{1}{2} \alpha m_\phi^2(M_\planck) \phi_0^2
\eeq

Decomposing $\phi$ as $\phi = \left( \phi_0 + s / \sqrt{2} \right) \exp \left[ ia / \left( \sqrt{2} \phi_0 \right) \right]$, the physical mass of saxion ($s$), denoted as $m_\mathrm{PQ}$, is 
\beq \label{mphi}
m_\mathrm{PQ}^2 
\equiv \frac{1}{2} \left. \frac{d^2 V}{d |\phi|^2} \right|_{\phi_0} 
\simeq \left. \frac{d m_\phi^2}{d \ln |\phi|} \right|_{\phi_0} 
\simeq 
\frac{1}{8 \pi^2} |\lambda_\Phi|^2 \left( m_\nu^2 + |A_\Phi|^2 \right)
\eeq
The masse of axon ($a$) at zero temperature is \cite{Amsler:2008zzb}
\beq \label{ma}
m_a \sim 6 \times 10^{-5} \eV \left( \frac{10^{12} \GeV}{f_a} \right) 
\eeq
where the axion coupling constant $f_a$ is defined as
\beq \label{fawrtphi0}
f_a \equiv \frac{\sqrt{2} \phi_0}{N}
\eeq
with $N=6$ the coefficient of $U(1)_\mathrm{PQ}$-QCD anomaly in our model in normalization of unit PQ charge for PQ field ($\phi$).
$f_a$ is lower-bounded by the cooling rate of SN 1987A \cite{Buckley:2007tm} and upper-bounded by over-closure limit of cold axions from misalignment and strings \cite{Amsler:2008zzb}.  
If there is no dilution after axion condensation from misalignment is formed, currently allowed window of $f_a$ is given by
\beq \label{axionwindow}
10^9 \GeV \lesssim f_a \lesssim 10^{12} \GeV
\eeq

The masse of axino ($\axino$) is generated at 1-loop in our model.
It is given by  \cite{Moxhay:1984am}
\beq \label{maxino}
m_\axino = \frac{1}{16 \pi^2} \lambda_\Phi^2 A_\Phi \simeq 0.6 \GeV \lambda_\Phi^2 \left( \frac{A_\Phi}{100 \GeV} \right)
\eeq
hence axino is the lightest supersymmetric particle(LSP) in our model.

The large vev of $\phi$, implied in \eq{axionwindow}, also makes $\nu$ become very heavy due to the Yukawa coupling associated with $\lambda_\Phi$ in \eq{W}, and integrating out $\nu$ in \eq{W} gives an effective low-energy superpotential which contains a left-handed neutrino mass term:
\beq \label{Weff}
W_\mathrm{eff} 
= \lambda_u Q H_u \bar{u} + \lambda_d Q H_d \bar{d} + \lambda_e L H_d \bar{e} + \lambda_\mu \frac{\phi_0^2 H_u H_d}{M_\planck} - \frac{1}{2} \frac{\lambda_\nu^2 \left( LH_u \right)^2}{\lambda_\Phi \phi_0}
\eeq
In order to provide right size of masses to left-handed neutrinos, we need
\beq \label{lambdaphicond}
\lambda_\Phi 
= \frac{\lambda_\nu^2 v^2 \sin^2 \beta}{m_{\nu_L} \phi_0} 
\simeq 0.33 \left( \frac{\lambda_\nu}{10^{-2}} \right)^2 \left( \frac{10^{-2} \eV}{m_{\nu_L}} \right) \left( \frac{10^{12} \GeV}{\phi_0} \right) \sin^2 \beta
\eeq
where $m_{\nu_L}$ is the left-handed neutrino mass, $v=174 \GeV$ is the vev of Higgs field and $\sin \beta \equiv v_u / v$ with $v_u$ the vev of up-type Higgs field ($H_u$).
Note that in \eq{Weff} MSSM $\mu$ parameter is given by
\beq
\mu = \lambda_\mu \frac{\phi_0^2}{M_\planck} = 4.2 \TeV \left( \frac{\lambda_\mu}{10^{-2}} \right) \left( \frac{\phi_0}{10^{12} \GeV} \right)^2
\eeq
Note also that using \eq{lambdaphicond}, we can re-express \eq{phi0} as 
\bea
\frac{m_\phi(M_\planck)}{\sqrt{m_\nu^2 + |A_\Phi|^2}} 
&\simeq& \frac{\lambda_\Phi}{2 \sqrt{2} \pi} \sqrt{- \frac{1}{2} - \ln \frac{\lambda_\Phi \phi_0}{M_\planck} }
\\ \label{m0overm}
&\simeq& 0.14 \left( \frac{\lambda_\nu}{10^{-2}} \right)^2 \left( \frac{10^{-2} \eV}{m_{\nu_L}} \right) \left( \frac{10^{12} \GeV}{\phi_0} \right) \sin^2 \beta
\eea
which provides a constraint on mass parameters.

As described before this, introduction of a gauge singlet chiral superfield allows natural realization of the PQ-symmetry to resolve the strong CP problem, and explains the origin of small mass of left-handed neutrino and MSSM $\mu$-parameter.

Our model is similar to the models of refs. \cite{Lazarides:1985bj} studied in refs. \cite{Choi:1996vz} 
and those of refs. \cite{Stewart:1996ai,Kim:2008yu}, but it differs from those models mainly by how symmetry breaking field is stabilized and/or how a right-handed neutrino obtains large mass for a see-saw mechanism.
In our model, instead of non-renormalizable higher order operator, the effect of field-dependent running of a mass-squared parameter is used to stabilize the symmetry breaking field whose vacuum expectation value provides a large mass to right-handed neutrinos.
Although it may be thought of a rather simple variation of the models, the difference makes our model simpler and maybe more natural with rather different cosmology which will be described in the next section.

\section{Cosmology}

As our ansatz, we assume only $LH_u$, $H_uH_d$ flat directions \cite{Casas:1995pd} and $\phi$ develop non-zero field values temporarily or eventually
\footnote{Although $H_uH_d$ may be stable near the origin as we are assuming here, dynamics of fields can lead non-zero field value of the flat direction.    
}.
This will be justified shortly in subsequent arguments.
Parametrizing $LH_u$ and $H_uH_d$ flat directions as
\beq
\begin{array}{ccccc}
L = \left( 0 , l \right)^T & , & H_u = \left( h_u , 0 \right)^T & , & H_d = \left( 0 , h_d \right)^T
\end{array}
\eeq
with the remaining $D$-term constraint
\beq
D = |h_u|^2 - |h_d|^2 - |l|^2 = 0
\eeq 
we find a potential from \eqs{W}{Vphi}
\bea
V
&=& m_L^2 |l|^2 + m_{H_u}^2 |h_u|^2 + m_{H_d}^2 |h_d|^2 + m_\phi(|\phi|) |\phi|^2 
\\
&& + \left( A_\mu \lambda_\mu \frac{\phi^2}{M_\planck} h_u h_d + \mathrm{c.c.} \right)
\\ \label{HighT-Fterm}
&& + \left| \lambda_\mu \frac{\phi^2}{M_\planck} h_u \right|^2 + \left| \lambda_\mu \frac{\phi^2}{M_\planck} h_d \right|^2 + \left| 2 \lambda_\mu \frac{\phi}{M_\planck} h_u h_d \right|^2 + \left| \lambda_\nu l h_u \right|^2
\\
&& + \frac{1}{2} g^2 \left( |h_u|^2 - |h_d|^2 - |l|^2 \right)
\eea  
where terms in the first to forth lines are soft SUSY-breaking mass terms, $A$-terms, $F$-terms and $D$-terms, respectively, and 
$g^2 = (g_1^2 + g_2^2) / 4$.
Based on this potential, in this section we will describe cosmology of our model including thermal inflation which may be indispensable to resolve moduli problem, baryogenesis and dark matter.

\subsection{Moduli problem and thermal inflation}

In string/M theories, that might be candidates of the fundamental theory, there are flat directions called moduli.
A modulus ($\varphi$) has interactions suppressed by Planck scale, and starts to oscillate coherently when expansion rate becomes comparable to the mass of the modulus, $m_\varphi$.
We assume that radiation is dominant at this time, then the abundance of the moduli (called Big-Bang moduli) is 
\beq \label{BBmoduli}
\left( \frac{n_\varphi}{s} \right)_\mathrm{BB}
\sim \left( \frac{M_\planck}{m_\varphi} \right)^{1/2}
\simeq 4.5 \times 10^7 \left( \frac{1 \TeV}{m_\varphi} \right)^{1/2}
\eeq
where $n_\varphi$ is the number density of moduli and $s$ is the entropy density.
The decay rate of moduli is given by
\beq
\Gamma_\varphi 
= \gamma_\varphi \frac{1}{8 \pi} \frac{m_\varphi^3}{M_\planck^2}
\simeq 6.9 \times 10^{-30} \GeV \gamma_\varphi \left( \frac{m_\varphi}{1 \TeV} \right)^3
\eeq
where $\gamma_\varphi \sim \mathcal{O}(1)$ is a numerical coefficient, hence moduli decay after BBN.
In order for successful BBN, the abundance of moduli when they decay is constrained to be \cite{Kawasaki:2004qu}
\beq \label{moduliBBNbound}
\frac{n_\varphi}{s} 
\lesssim \frac{10^{-14} \GeV}{m_\varphi}
\simeq 10^{-17} \left( \frac{1 \TeV}{m_\varphi} \right)
\eeq
Therefore, we need a dilution more than $\mathcal{O}(10^{24})$.

Thermal inflation is the most compelling solution to this moduli problem.
Moduli may start to oscillate coherently while $\phi$ is held around the origin due to the interaction with $\nu$ which is in very hot thermal bath.
Thermal inflation begins when the potential energy density at the origin becomes dominant over the energy density of moduli at a temperature given by 
\beq \label{Tb}
T_\mathrm{b} \sim \rho_\varphi^{1/4} \left( \frac{V_0}{\rho_\varphi} \right)^{1/3} \sim \left( \frac{V_0^2}{m_\varphi M_\planck} \right)^{1/6}
\eeq
where we have used $\rho_\varphi \sim m_\varphi^2 M_\planck^2$ as the initial energy density of moduli coherent oscillation. 
Thermal inflation ends as $\phi$ starts to roll out when temperature drops to the critical temperature
\beq \label{Tc}
T_\mathrm{c} \sim \sqrt{|m_\phi^2(|\phi| \sim 0)|} \sim \left| \ln \frac{m_\mathrm{soft}}{\lambda_\Phi \phi_0} \right|^{1/2} m_\mathrm{PQ}
\eeq
where $m_\mathrm{soft} \sim 1 \TeV$ is the scale of soft SUSY-breaking.

The coherent oscillation of $\phi$ after thermal inflation eventually decays with rate
\beq
\Gamma_\phi \simeq \Gamma_{\phi \to a a} + \Gamma_{\phi \to \mathrm{SM}}
\eeq
where 
\beq \label{Gammaphitoaa}
\Gamma_{\phi \to a a} = \frac{1}{64 \pi} \frac{m_\mathrm{PQ}^3}{\phi_0^2}
\eeq
is the partial decay rates of $\phi$ to axions.
$\Gamma_{\phi \to \mathrm{SM}}$ is the partial decay rates of $\phi$ to SM particles.
It is mainly due to the mixing between $\phi$ and Higgs fields, induced by the term $\lambda_\mu \Phi^2 H_u H_d / M_\planck$.
Since $m_\mathrm{PQ} \sim \mathcal{O}(10 - 100) \GeV$ for $\lambda_\Phi \sim \mathcal{O}(1)$ from \eq{mphi}, $\Gamma_{\phi \to \mathrm{SM}}$ is expected to be dominated by the decay to bottom quarks and given by \cite{Kim:2008yu}   
\beq \label{Gammaphitosm}
\Gamma_{\phi\to {\rm SM}}
\simeq 
\frac{1}{4 \pi} \left| \frac{m_A^2 - |B|^2}{m_A^2} \right|^2
\left( \frac{|\mu|^4}{m_\mathrm{PQ} \phi_0^2} \right)
\left[ \fn{f}{ \frac{m_h^2}{m_\mathrm{PQ}^2} } \right]
\eeq
where $m_A$ is the mass of CP-odd neutral Higgs particle, $B = A_\mu$ in our model, 
\beq \label{fx}
\fn{f}{x} = \frac{\varepsilon x}{(1-x)^2} \left( 1 - \frac{\varepsilon x}{3} \right)^\frac{3}{2}
\eeq
and
\beq \label{varepsilon}
\varepsilon \sim \frac{12 m_b^2}{m_h^2} \sim 0.02
\eeq
with $m_b$ and $m_h$ the masses of bottom quark and light neutral Higgs particle, respectively.
Successful BBN limits energy contributions of relativistic non-SM particles, so it requires $\Gamma_{\phi \to aa} / \Gamma_{\phi \to \mathrm{SM}} \lesssim 0.3$ \cite{Kim:2008yu} which is easily satisfied for $m_\mathrm{PQ} \sim \mathcal{O}(10 - 100) \GeV$.
From \eqs{Gammaphitoaa}{Gammaphitosm}, the decay temperature of flaton ($\phi$) is given by 
\bea \label{Td}
T_\mathrm{d}
&\equiv& \left( \frac{\pi^2}{15} g_*(T_\mathrm{d}) \right)^{-1/4} \left( \Gamma_{\phi \to \mathrm{SM}} \Gamma_\phi \right)^{1/4} M_\planck^{1/2}
\\ \label{Td1}
& \simeq &
\left( \frac{5}{8 \pi^4 \fn{g_*}{T_\mathrm{d}}} \right)^\frac{1}{4}
\left| 1 - \frac{|B|^2}{m_A^2} \right|
\frac{|\mu|^2}{m_\mathrm{PQ}^{1/2} \phi_0}
\left[ \fn{f}{ \frac{m_h^2}{m_\mathrm{PQ}^2} } \right]^\frac{1}{2}
\\ \label{Td2}
&\simeq& 26 \GeV
\left| 1 - \frac{|B|^2}{m_A^2} \right|
\left( \frac{10^{12} \GeV}{\phi_0} \right)
\left( \frac{|\mu|}{1 \TeV} \right)^2
\left( \frac{30 \GeV}{m_\mathrm{PQ}} \right)^\frac{1}{2}
\left[ \fn{f}{ \frac{m_h^2}{m_\mathrm{PQ}^2} } \right]^\frac{1}{2}
\eea
where we have used $g_*(T_\mathrm{d})=100$ in the last line.
\begin{figure}[t]
\centering
\includegraphics[width=0.75\textwidth]{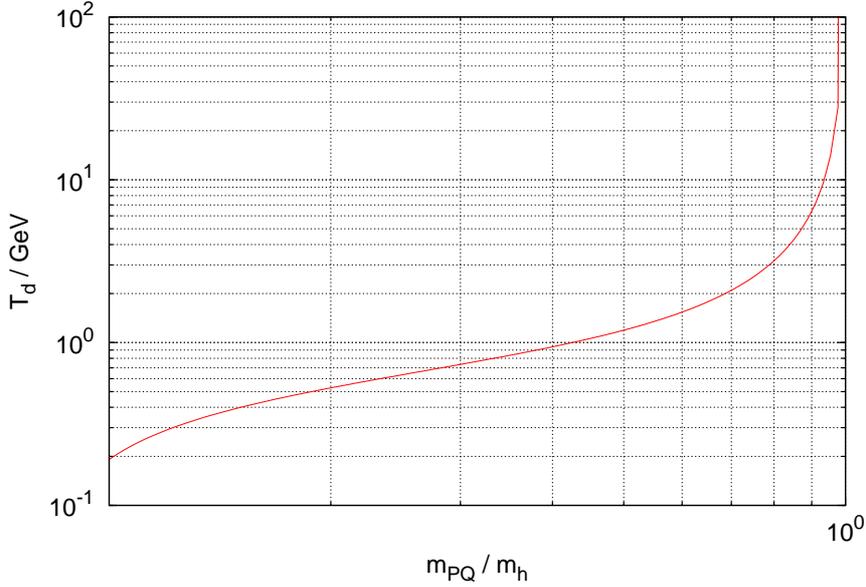}
\caption{ \label{fig:Td}
Flaton decay temperature versus flaton mass: $T_\mathrm{d}$ versus $m_\mathrm{PQ} / m_h$ for $\phi_0 = 10^{12} \GeV$, $|\mu| = 10^3 \GeV$, $m_h = 125 \GeV$ and $m_A = 2|B|$.
}
\end{figure}
Fig. \ref{fig:Td} shows $T_\mathrm{d}$ as a function of $m_\mathrm{PQ}/m_h$.

The $e$-foldings of thermal inflation is, from \eqs{Tb}{Tc}, 
\bea \label{phiefolds}
N_\phi
&=& \ln \frac{T_\mathrm{b}}{T_\mathrm{c}}
\\
&\sim& 7.6 + \frac{1}{6} \ln \left[ \left( \frac{\phi_0}{10^{12} \GeV} \right)^4 \left( \frac{30 \GeV}{m_\mathrm{PQ}} \right)^2 \left( \frac{1 \TeV}{m_\varphi} \right) \right]
\nonumber \\ 
&& \phantom{7.2} - \frac{1}{2} \ln \left| \ln \left( \frac{m_\mathrm{soft}}{1 \TeV} \right) - \ln \left( \frac{|\lambda_\Phi \phi_0|}{10^{12} \GeV} \right) \right|
\eea  
and the dilution factor due to the entropy release in the decay of $\phi$ is 
\bea \label{phidfactor}
\Delta_\phi
&\sim& \frac{V_0}{T_\mathrm{c}^3 T_\mathrm{d}}
\\
&\sim& 2 \times 10^{21}  \left( \frac{\phi_0}{10^{12} \GeV} \right)^2 \left( \frac{30 \GeV}{m_\mathrm{PQ}} \right) \left( \frac{10^{-1} \GeV}{T_\mathrm{d}} \right) \left| \ln \left( \frac{10^9 m_\mathrm{soft}}{|\lambda_\Phi \phi_0|} \right) \right|^{-3/2}
\eea
This is little bit small to dilute out moduli produced before thermal inflation to a safe level.

Fortunately, there is additional source of dilution.
We assume
\beq
T_\mathrm{c} < T_{LH_u}
\eeq
with
\beq
T_{LH_u}^2 \sim m_{LH_u}^2 \equiv - \frac{1}{2} \left( m_L^2 + m_{H_u}^2 \right)
\eeq
Then, as temperature drops below $T_{LH_u}$, $LH_u$ which is expected to be held near the origin at high temperature rolls out from the origin before flaton is destabilized.
It is stabilized by the $F$-term potential, $\left| \lambda_\nu l h_u \right|^2$ in \eq{HighT-Fterm} at
\beq
|l_*| \simeq |h_{u*}| \sim m_{LH_u} / |\lambda_\nu|
\eeq
The initial energy density of the coherent oscillation of $LH_u$ is 
\beq
V_{LH_u} \sim m_{LH_u}^2 |l_*|^2 \sim m_{LH_u}^4 / |\lambda_\nu|^2
\eeq
If this energy can be quickly dumped into radiation we would get an additional dilution to \eq{phidfactor}. 

The condensation of $LH_u$ decays initially through prheating \cite{Kofman:1994rk} and eventually through perturbative decays.
For $\lambda_\nu \gtrsim 10^{-2}$, the perturbative decay of the condensation is dominated by the coupling $\lambda_\nu$ with a partial decay rate given by 
\footnote{
If a coupling of $LH_u$ to MSSM particles is larger than $\lambda_\nu$, the particles becomes heavier than $LH_u$, hence the decay of $LH_u$ is kinematically forbidden. 
}
\beq
\Gamma_{LH_u} \sim \frac{1}{8 \pi} |\lambda_\nu|^2 m_{LH_u}
\eeq
Using \eqss{V0}{alpha}{lambdaphicond}, one finds
\beq
\frac{\Gamma_{LH_u}}{H_\mathrm{TI}} \sim \frac{280}{\sin^2 \beta} \left( \frac{m_{\nu_L}}{10^{-2} \eV} \right) \left( \frac{m_{LH_u}}{m_\nu} \right)
\eeq 
where $H_\mathrm{TI}$ is the Hubble expansion rate during thermal inflation, hence the coherent oscillation of $LH_u$ decays within a Hubble time.

It is expected that preheating is at best dominant for modes $k \sim m_{LH_u}$ and order one fractional energy density of $LH_u$ is still in the form of condensation, hence the Universe is slightly reheated to a temperature 
\beq \label{Textension}
T_* \sim V_{LH_u}^{1/4} \sim |\lambda_\nu|^{-1/2} m_{LH_u}
\eeq
before flaton is destabilized.
Thermal inflation is then extended by $e$-foldings given by
\beq \label{LHuefolds}
N_{LH_u} = \ln \frac{T_*}{T_{LH_u}} \sim - \frac{1}{2} \ln |\lambda_\nu|
\eeq
and, using \eq{Textension}, the dilution factor due to the entropy release in the decay of $LH_u$ is 
\beq \label{LHudfactor}
\Delta_{LH_u} \sim \frac{V_{LH_u}^{3/4}}{m_{LH_u}^3} \sim |\lambda_\nu|^{-3/2}
\eeq
Therefore, from Eqs. (\ref{BBmoduli}), (\ref{phidfactor}), (\ref{LHudfactor}) and (\ref{lambdaphicond}) the abundance of Big-Bang moduli when they decay is 
\bea \nonumber
\frac{\left( n/s \right)_\mathrm{BB}}{\Delta_{LH_u} \Delta_\phi} 
&\sim& \left( \frac{M_\planck}{m_\varphi} \right)^{1/2} \frac{T_\mathrm{c}^3 T_\mathrm{d}}{V_0} |\lambda_\nu|^{3/2}
\\
&\sim& 2 \times 10^{-17} \left( \frac{1 \TeV}{m_\varphi} \right)^{1/2} \left( \frac{|\lambda_\nu|}{10^{-2}} \right)^{3/2}
\\ \label{BBmoduliatdecaytime}
&& \times  \left( \frac{10^{12} \GeV}{\phi_0} \right)^2 \left( \frac{m_\mathrm{PQ}}{30 \GeV} \right) \left( \frac{T_\mathrm{d}}{10^{-1} \GeV} \right) \left| \ln \left( \frac{10^9 m_\mathrm{soft}}{|\lambda_\Phi \phi_0|} \right) \right|^{3/2}
\eea
which can manage to satisfy the bound from observation in \eq{moduliBBNbound}.
Thus, Big-Bang moduli can be diluted out to a safe level.

Moduli are also produced after thermal inflation due to the finite energy density of thermal inflation.
The abundance when they are produced is
\beq \label{TImoduli}
\left( \frac{n_\varphi}{s} \right)_\mathrm{TI}
\sim \frac{V_0^2}{T_\mathrm{c}^3 m_\varphi^3 M_\planck^2}
\eeq
Using \eqs{phidfactor}{V0}, the abundance when they decay is
\bea \label{TImoduliatdecaytime}
\frac{ \left( n_\varphi / s \right)_\mathrm{TI}}{\Delta_\phi} 
&\sim& \frac{V_0 T_\mathrm{d}}{m_\varphi^3 M_\planck^2}
\\
&\sim& 8 \times 10^{-21} \left( \frac{m_\mathrm{PQ}}{30 \GeV} \right)^2 \left( \frac{\phi_0}{10^{12} \GeV} \right)^2  \left( \frac{T_\mathrm{d}}{100 \MeV} \right) \left( \frac{1 \TeV}{m_\varphi} \right)^3 
\eea 
Compared to Big-Bang moduli of \eq{BBmoduliatdecaytime}, it is negligible, hence it is harmless as long as Big-Bang moduli is diluted to a safe level.

Apart from the dilution of unwanted relics, thermal inflation has a very interesting aspect. 
It wipes out pre-existing gravitational wave backgrounds and generates its own one which may be detected at BBO or DECIGO type experiment \cite{Easther:2008sx}.
Since the peak frequency and amplitude depend on the details of model parameters and physics of phase transition, careful investigation may be necessary to see if the wave in our model is really detectable.

\subsection{Baryogenesis}
As $\phi$ rolls out below $T_\mathrm{c}$, thermal inflation ends and $\nu$ becomes very heavy.
By integrating out $\nu$ in \eq{W}, one finds an effective superpotential
\beq \label{Weff1}
W 
= \lambda_u Q H_u \bar{u} + \lambda_d Q H_d \bar{d} + \lambda_e L H_d \bar{e} 
+ \lambda_\mu \frac{\phi^2 H_u H_d}{M_\planck} - \frac{1}{2} \frac{\lambda_\nu^2 \left( LH_u \right)^2}{\lambda_\Phi \phi}
\eeq
which gives a potential  
\bea 
V \label{massterm}
&=& m_L^2 |l|^2 + m_{H_u}^2 |h_u|^2 + m_{H_d}^2 |h_d|^2 + m_\phi^2(\phi) |\phi|^2 
\\ \label{Aterm}
&& + \left( A_\mu \lambda_\mu \frac{\phi^2 h_u h_d}{M_\planck} - \frac{1}{2} A_\nu \frac{\lambda_\nu^2 \left( l h_u \right)^2}{\lambda_\Phi \phi} + \mathrm{c.c.} \right)
\\ \label{Fterm}
&& + \left| \lambda_\mu \frac{\phi^2 h_d}{M_\planck} - \frac{\lambda_\nu^2 l^2 h_u}{\lambda_\Phi \phi} \right|
+ \left| \lambda_\mu \frac{\phi^2 h_u}{M_\planck} \right|
+ \left| 2 \lambda_\mu \frac{\phi h_u h_d}{M_\planck} + \frac{1}{2} \frac{\lambda_\nu^2 l^2 h_u^2}{\lambda_\Phi \phi^2} \right|
\\
&& + \frac{1}{2} g^2 \left( |h_u|^2 - |h_d|^2 - |l|^2 \right)^2
\eea
for $\phi \gg m_\mathrm{soft}$.

As $\phi$ becomes large, $LH_u$ is shifted to a larger value
\beq \label{l}
|l|^2 \simeq |h_u|^2 \simeq \frac{\left| \lambda_\Phi \phi \right| \left( \sqrt{12 m_{LH_u}^2 + |A_\nu|^2} + |A_\nu| \right)}{6 \left| \lambda_\nu^2 \right|}
\eeq
For $m_\mathrm{soft} \ll \phi \ll \phi_0$, the phase of $LH_u$ is determined by the $A_\nu$-term in \eq{Aterm}.
As $\phi$ reaches its vev, MSSM $\mu$-parameter is reproduced, hence $LH_u$ is lifted up and brought back into the origin
\footnote{
In order to avoid possible local minimum along $LH_u$, we need $m_L^2 + m_{H_u}^2 + |\mu|^2 > |A_\nu|^2 /6$, otherwise $LH_u$ may be trapped there that is disastrous.
}.
Simultaneously, $H_uH_d$ becomes large \footnote{Large $H_uH_d$ holds dangerous quark and lepton flat directions near the origin, hence considering only $LH_u$, $H_uH_d$ and $\phi$ is justified.} due to the cross term of the third term in \eq{Fterm}, and the cross term of the first term in the same equation provides angular kick to $LH_u$.
This is nothing but an Affleck-Dine (AD) mechanism \cite{Affleck:1984fy} to generate charge asymmetry.

The generated lepton asymmetry is expected to be conserved due to the rapid preheating of the AD fields \cite{Kim:2008yu}.
The oscillating AD fields decay earlier than flaton, reheat the Universe partially and activate sphaleron process \cite{Kuzmin:1985mm} in which lepton asymmetry is converted to baryon asymmetry.
Eventually, flaton decays and reheats the Universe in order for successful BBN.
There is huge amount of entropy release in the flaton decay.
The baryon asymmetry at present is then given by
\beq \label{baryonasymmetry}
\frac{n_B}{s} \sim \frac{n_B}{n_\phi} \frac{T_\mathrm{d}}{m_\mathrm{PQ}} \sim \frac{n_L}{n_\mathrm{AD}} \frac{n_\mathrm{AD}}{n_\phi} \frac{T_\mathrm{d}}{m_\mathrm{PQ}} \sim \frac{n_L}{n_\mathrm{AD}} \frac{m_{LH_u}}{m_\mathrm{PQ}} \left( \frac{|l_0|}{\phi_0} \right)^2 \frac{T_\mathrm{d}}{m_\mathrm{PQ}}
\eeq
where $n_\phi$, $n_L$ and $n_\mathrm{AD}$ are number densities of $\phi$, lepton asymmetry and AD field respectively, and $l_0$ is the value of $l$ when $\phi$ reaches $\phi_0$ from the origin.
From the experience of a similar model of Ref. \cite{Kim:2008yu}, we expect
\beq
\frac{n_L}{n_\mathrm{AD}} \sim \mathcal{O}(10^{-3} \textrm{ to } 10^{-2})
\eeq
and from \eqs{l}{lambdaphicond}
\beq
|l_0| \sim 10^9 \GeV \sqrt{\left( \frac{m_{LH_u}}{1 \TeV} \right) \left( \frac{10^{-2} \eV}{m_{\nu_L}} \right)}
\eeq
hence we expect
\bea
\frac{n_B}{s} &\sim& 10^{-10} \left( \frac{n_L / n_\mathrm{AD}}{10^{-3}} \right) 
\nonumber \\
&& \times 
\left( \frac{m_{LH_u}}{1 \TeV} \right)^2 \left( \frac{30 \GeV}{m_\mathrm{PQ}} \right)^2 \left( \frac{10^{-2} \eV}{m_{\nu_L}} \right) \left( \frac{10^{12} \GeV}{\phi_0} \right)^2 \left( \frac{T_\mathrm{d}}{100 \MeV} \right)  
\label{nBs}
\eea
Note that the model parameters to resolve moduli problem give right order of the baryon asymmetry which can match the present observation.

\subsection{Dark matter}

In our model, axinos and axions are expected to be the main components of dark matter at present.
These dark matter components can be cold, warm or even hot, depending on their masses and how they are produced. 
In this subsection, we will derive cosmological constraints from the dark matter.

\paragraph{Axinos produced in the flaton decay}
As the LSP, axinos are produced mainly in the decay of flaton with a rate given by \cite{Kim:2008yu}
\beq \label{phitoaxino}
\Gamma_{\phi \to \axino \axino} = \frac{\alpha_\axino^2 m_\axino^2 m_\phi}{32 \pi \phi_0^2}
\eeq
where $\alpha_\axino \equiv \left( d \ln m_\axino \right) / \left( d \ln |\phi| \right)$.
In our model, from \eqss{maxino}{aphiRGE}{lambdaphiRGE},
\beq
\alpha_\axino \simeq \frac{5}{8 \pi^2} |\lambda_\Phi|^2
\eeq
The axino number density from the flaton decay is 
\beq \label{naxino}
n_\axino = \frac{2 \Gamma_{\phi \to \axino \axino}}{m_\phi a(t)^3} \int_0^t a(t')^3 \rho_\phi dt'
\eeq
where $a$ is the scale factor and $\rho_\phi$ is the energy density of $\phi$.
Using \eqs{Td}{naxino}, one find \cite{Kim:2008yu}
\beq
\frac{n_\axino}{s} = \frac{2.2 T_\mathrm{d} \Gamma_{\phi \to \axino \axino}}{m_\phi \Gamma_\mathrm{SM}}
\eeq
From \eqs{Td}{phitoaxino}, the current axino abundance is 
\bea
\Omega_\axino 
&\simeq& 5.6 \times 10^8 \left( \frac{m_\axino}{1 \GeV} \right) \frac{n_\axino}{s}
\\
&\simeq& 0.36 \frac{\Gamma_\phi^{1/2}}{\Gamma_\mathrm{SM}^{1/2}} \left( \frac{10}{g_*^{1/2}(T_\mathrm{d})} \right) \left( \frac{\alpha_\axino}{10^{-1}} \right)^2 \left( \frac{m_\axino}{1 \GeV} \right)^3 \left( \frac{100 \MeV}{T_\mathrm{d}} \right) \left( \frac{10^{12} \GeV}{\phi_0} \right)^2
\eea
Therefore $\Omega_\axino \leq \Omega_\mathrm{CDM} \simeq 0.25$ requires
\beq \label{axinomassbound}
m_\axino \lesssim 0.9 \GeV \left[
\frac{\Gamma_\mathrm{SM}^{1/2}}{\Gamma_\phi^{1/2}}
\left( \frac{\fn{g_*^{1/2}}{T_\mathrm{d}}}{10} \right)
\left( \frac{T_\mathrm{d}}{100 \MeV} \right)
\left( \frac{10^{-1}}{\alpha_\axino} \vphantom{\frac{T_\mathrm{d}}{1 \GeV}} \right)^2
\left( \frac{\phi_0}{10^{12} \GeV} \right)^2
\right]^\frac{1}{3}
\eeq
which can be easily satisfied for $\lambda_\Phi \lesssim 1$ from \eq{axinomassbound}.

\paragraph{Axinos in the decay of thermally generated NLSP}

If the next-to-lightest supersymmetric particle (NLSP) denoted as $\chi$ is of neutralino type, the coupling $\lambda_\mu \phi^2 H_u H_d / M_\planck$ in \eq{W} leads to the decay of $\chi$ to axino via various channels with rate \cite{Covi:1999ty}
\bea
\label{gammahiggsino}
\Gamma_{\chi \to \axino} &=& \frac{C_{\axino}}{16\pi}
\frac{m^3_\chi}{\varphi_0^2},
\eea
where $C_\axino \sim 1$ may contain a factor of $m^2_Z/m^2_\chi$ and
we have neglected the masses of decay products.

The thermal bath generates $\chi$ with the number density
\bea
n_\chi &=& \frac{1}{\pi^2} \int_0^\infty d k \frac{k^2}{\exp\sqrt{\frac{k^2+m_\chi^2}{T^2}} + 1},
\eea
and they subsequently decay into axinos.
The late time axino abundance is thus estimated as
\bea \label{ndecay}
n_\axino 
&=& \frac{\Gamma_{\chi \to \axino}}{a(t)^3} \int_0^t a(t')^3 n_\chi(t') d t'  
\eea
whose numerical solution results in \cite{Kim:2008yu}
\bea \label{naxino}
\frac{n_\axino}{s} 
&=& g^{-1/4}_\ast(T_{\rm d}) g^{-5/4}_\ast(T_\chi) \left(\frac{\Gamma_{\phi \to \mathrm{SM}}}{\Gamma_\phi}\right)^{1/2} \frac{\Gamma_{\chi \to \axino} M_\planck}{m^2_\chi} F_\axino(x)
\eea
where $F_\axino(x)$ is approximated as
\bea
F_\axino(x) &\sim&
\left\{
\begin{array}{ccc}
5.3 x^7 & {\rm for} & x \ll 1
\\
5.4 & {\rm for} & x \gg 1
\end{array}
\right.
\eea
with
\bea
x &=& \frac{2}{3} \left(\frac{g_\ast(T_{\rm d})}{g_\ast(T_\chi)}\right)^{1/4} \frac{T_{\rm d}}{T_\chi}
\eea
and $T_\chi\simeq 2m_\chi/21$ being the temperature at which the axino production rate is maximized.
From (\ref{gammahiggsino}) and (\ref{naxino}), the current abundance of axino dark matter is obtained
\bea \label{omegaaxinoflatoncon}
\Omega_\axino 
&\simeq& 5.6\times 10^8 \left(\frac{m_\axino}{1{\rm GeV}}\right) \frac{n_\axino}{s} 
\\
&\simeq&
2.7 C_\axino \Big( \frac{10^3}{g^{1/4}_\ast(T_{\rm d})g^{5/4}_\ast(T_\chi)} \Big) \left(\frac{m_\chi}{10^2{\rm GeV}}\right) \left(\frac{m_\axino}{1{\rm GeV}}\right) \left(\frac{10^{12}{\rm GeV}}{\phi_0}\right)^2 F_\axino(x)
\nonumber \\
\eea
where we have used $\Gamma_\phi \simeq \Gamma_{\phi \to {\rm SM}}$.
Therefore, for $x\ll 1$, one finds
\bea \label{approxomeganlsp}
\Omega_\axino 
\simeq 1.2\times 10^7\, C_\axino
\left(\frac{T_{\rm d}}{m_\chi}\right)^7
\Big(\frac{10^3 g^{3/2}_\ast(T_{\rm d})}{g^3_\ast(T_\chi)}\Big)
\left(\frac{m_\chi}{10^2{\rm GeV}}\right)
\left(\frac{m_\axino}{1{\rm GeV}}\right)
\left(\frac{10^{12}{\rm GeV}}{\phi_0}\right)^2
\eea
which does not exceed $\Omega_{\rm CDM}$ if the flaton decay temperature satisfies
\bea \label{Tdbound}
T_{\rm d} 
&\lesssim& 7.5 \GeV \left( \frac{m_\chi}{100 \GeV} \right) \left(\frac{g_\ast(T_\chi)}{g_\ast(T_{\rm d})}\right)^{1/4}
\nonumber \\
&&
\times
\left[C^{-1}_\axino
\Big(\frac{g^{1/4}_\ast(T_{\rm d})g^{5/4}_\ast}{10^3}\Big)
\left(\frac{10^2{\rm GeV}}{m_\chi}\right)
\left(\frac{1{\rm GeV}}{m_\axino}\right)
\left(\frac{\phi_0}{10^{12}{\rm GeV}}\right)^2 \right]^{1/7}
\eea
Note that in order to resolve moduli problem (See \eq{BBmoduliatdecaytime}) and explain baryon asymmetry of the Universe (See \eq{baryonasymmetry}) we need $T_\mathrm{d} \sim \mathcal{O}(100) \MeV$, and in that case \eq{Tdbound} is satisfied automatically.
Since $T_{\rm d}$ is expected to be much less than the freeze-out temperature of $\chi$, which is about $m_\chi/20$ \cite{Amsler:2008zzb}, in order to avoid direct production of $\chi$ from the flaton decay, we may require $m_\phi< 2m_\chi$, which seems to be typical in our model.

Axinos will also be produced by the decay of $\chi$ after they freeze out.
However, the standard Big-Bang neutralino freeze-out abundance is good match to the dark matter abundance, our freeze-out abundance of $\chi$ will typically be less than the standard abundance, and $m_\axino\ll m_\chi$, therefore the axino abundance generated after the freeze-out should be safe.

\paragraph{Axions from misalignment and strings}
The current abundance of axions produced in these ways is given by \cite{Amsler:2008zzb}
\beq \label{Omegaa}
\Omega_a = \frac{0.56}{\Delta_a} \left( \frac{f_a}{10^{12} \GeV} \right)^{1.167}
\eeq
where $\Delta_a$ is the dilution factor of axion misalignment due to the late time entropy release of thermal inflation.
For $T_\mathrm{d} \ll 1.3 \GeV$ \cite{Kim:2008yu},
\beq \label{Deltaa}
\Delta_a \sim \left( \frac{1.3 \GeV}{T_\mathrm{d}} \right)^{1.96}
\eeq
hence for $\Omega_a^\mathrm{cold} \leq \Omega_\mathrm{CDM} \simeq 0.25$ we need 
\beq
T_\mathrm{d} \lesssim 2.0 \GeV \left( \frac{10^{12} \GeV}{\phi_0} \right)^{1.167/1.96}
\eeq
Since we expect $T_\mathrm{d} \lesssim \mathcal{O}(100) \MeV$, these axions are likely to be a sub-dominant contribution to cold dark matter at present.

\paragraph{Axions in the flaton decay}

Axion can be warm or hot when they are produced in the decay of flaton.
The constraint on hot dark matter comes from CMBR and structure formation \cite{Bashinsky:2003tk,Hannestad:2003ye}.
Currently allowed hot dark matter fractional contribution to the present critical density is $\Omega_{HDM} \lesssim 10^{-2}$ \cite{Amsler:2008zzb}.
The constraint may be more stringent as suggested from the analysis of the early re-ionization of the Universe at high redshift \cite{Jedamzik:2005sx}.
Taking into account the recent analysis of WMAP $5$-year data \cite{Dunkley:2008ie}, the allowed warm/hot dark matter fractional
contribution to the present critical density is likely to be
\bea \label{WHDMconstraint}
\Omega_{\rm WHDM} &\lesssim& 10^{-3}.
\eea
We will take this as the upper bound on the fractional energy density of our warm/hot dark matter.

Axions produced by the flaton decay have a current momentum
\bea \label{pa}
p_a &=& \frac{a}{a_0}\frac{m_\phi}{2},
\eea
where $a$ is the scale factor at the time they were created and $a_0$ is the scale factor now.
Whereas, the current momentum of an axion produced at $t_{\rm d} = \Gamma_\phi^{-1}$ is
\bea
p_{\rm d} 
&=& \frac{a_{\rm d}}{a_0} \frac{m_\phi}{2}
\\
&=& \frac{S_{\rm d}^{1/3} g_{\ast S}^{1/3}(T_0)T_0}{S_{\rm f}^{1/3} g_{\ast S}^{1/3}(T_{\rm d})T_{\rm d}} \frac{m_\phi}{2}
\\
&\simeq& 2.06 \times 10^{-2} \eV \left( \frac{10}{g_\ast(T_{\rm d})} \right)^{1/3} \left( \frac{100 \MeV}{T_\mathrm{d}} \right) \left( \frac{m_\phi}{30 \GeV} \right)
\eea
where $S_\mathrm{d}$ and $S_\mathrm{f}$ are respectively the total entropy at decay time $t_\mathrm{d}$ and present,
so it would be highly relativistic now.
The current number density spectrum is given by
\bea
p_a \frac{dn_a^{\rm hot}}{dp_a} 
&=&
\left( \frac{a}{a_0} \right)^3 \frac{2 \rho_\phi}{m_\phi} \frac{\Gamma_{\phi \to a a}}{H}
= \frac{16  p_{\rm d}^3}{m^4_\phi} \frac{\Gamma_{\phi \to a a} p_a^3}{H p_{\rm d}^3}\rho_\phi,
\eea
which may provide an observational test of our model in the future.
The energy density of the axions is
\bea \label{rhoaflaton}
\frac{\rho_a^{\rm hot}}{\rho_{\rm SM}}
&=&
\frac{g_\ast(T_{\rm d}) g_{\ast S}^{4/3}(T)}{g_\ast(T) g^{4/3}_{\ast S}(T_{\rm d})} \frac{\Gamma_{\phi \to a a}}{\Gamma_{\phi\to {\rm SM}}}
\eea
Therefore, assuming that the hot axions are still relativistic now, their current energy density is estimated as\footnote{The energy density of thermally produced axions is $\Omega_a \sim m_a/131{\rm eV}$, hence it will be subdominant \cite{KT}.}
\bea \label{omegaaflaton}
\Omega_a^\mathrm{hot} 
&\simeq& 4.3 \times 10^{-5} \left( \frac{10}{g_\ast(T_{\rm d})} \right)^\frac{1}{3} \frac{\Gamma_{\phi\to a a} }{\Gamma_{\phi\to {\rm SM}}}
\eea

\section{Conclusion}

In this paper, we have proposed maybe the simplest extension of MSSM by introducing just one additional gauge singlet chiral superfield and right-handed neutrino superfield ($\nu$) to MSSM without any ad-hoc mass parameter except soft SUSY-breaking ones:
\beq
W 
= \lambda_u Q H_u \bar{u} + \lambda_d Q H_d \bar{d} + \lambda_e L H_d \bar{e} + \lambda_\nu L H_u \nu 
+ \lambda_\mu \frac{\Phi^2 H_u H_d}{M_\planck} + \frac{1}{2} \lambda_\Phi \Phi \nu^2
\eeq
with an assumption for the soft mas-squared parameters of $L$ and $H_u$   
\beq
m_L^2 + m_{H_u}^2 < 0
\eeq
and surely $m_L^2 + m_{H_u}^2 + |\mu|^2 > 0$ around electroweak scale.
This model can realizes Peccei-Quinn symmetry to resolve strong CP problem and explains the origin of left-handed neutrino mass and $\mu$-parameter of MSSM.
In addition, the model realizes thermal inflation followed by Affleck-Dine type leptogenesis and provides dark matter which matches well to observation in a remarkably consistent way. 

At low energy, the scalar component of the gauge singlet field, $\phi$ has negative mass-squared around the origin due to strong Yukawa coupling to $\nu$, hence develops non-zero vacuum expectation value.
In the absence of self-interactions, the field can be stabilized radiatively around intermediate scale, so it can be identified as the Pecci-Quinn field which breaks the Pecci-Quinn symmetry spontaneously.
The vacuum expectation value of $\phi$ make $\nu$ heavy, hence integrating out $\nu$ at low energy generates the mass term of left-handed neutrino through seesaw mechanism.
The vev of $\phi$ also generates MSSM $\mu$-parameter.

Cosmologically, this model gives rise to thermal inflation while $\phi$ is held near the origin due to large thermal effect at high temperature.
Notably, the non-trivial assumption of $m_L^2 + m_{H_u}^2 < 0$ allows $LH_u$ flat direction to be destabilized before the end of thermal inflation.
This causes an extension of thermal inflation by few $e$-foldings.
Although it is small, the additional $e$-foldings are crucial to dilute out moduli to a safe level. 
Thermal inflation ends as $\phi$ is destabilized at a critical temperature.
As $\phi$ reaches its vacuum expectation value, MSSM $\mu$-parameter is reproduced, leading Affleck-Dine leptogenesis involving $LH_u$ and $H_uH_d$ flat directions.
The oscillation of Affleck-Dine fields is expected to be quickly damped due to a rapid preheating induced by oscillation and preheating of flaton field $\phi$, so the generated lepton asymmetry can be conserved.

The eventual decay of flaton reheats the Universe with a temperature $T_\mathrm{d} \sim \mathrm{O}(100) \MeV$, releasing huge amount of entropy.
By its own, the entropy release in the decay of flaton is not enough to resolve moduli problem.
But the decay of $LH_u$ condensation before the end of thermal inflation provides an additional dilution.
As the result, moduli can be dilute out to a safe level.
Interestingly axion coupling constant and flaton decay temperature are  tightly constrained to be $f_a \sim 10^{12} \GeV$ and $T_\mathrm{d} \sim 100 \MeV$, respectively.
Remarkably, model parameters in this case are likely to give naturally right amount of baryon asymmetry at present. 

The main component of dark matter in our model is expected to be the axino whose mass is nearly fixed to be about $1 \GeV$ to match cold dark matter abundance at present.
Axions from misalignment and strings also contribute to cold dark matter, but they are sub-dominant. 
Axions from the decay of flaton are sub-dominant too, but they are expected to be hot and highly relativistic at present in our model.
The current number density spectrum of the hot axions may provide experimental and/or observational tests of our model in the future.       
The current number density spectrum of the hot axions may provide an observational test of our model in the future.

\subsection*{Acknowledgements}
WIP thanks to Ewan D. Stewart for valuable comments.
WIP is supported by the KRF grant (KRF-2008-314-C00064), KOSEF grants (KOSEF 303-2007-2-C000164 and No. 2009-0077503) and Brain Korea 21 projects funded by the Korean Government.

\appendix

\section{Renormalization group equations}
The relevant part of superpotential of \eq{W} for the running of soft SUSY-breaking parameters associated with $\phi$ and $\nu$ is 
\beq
W_\mathrm{part} = \lambda_\nu L H_u \nu + \frac{1}{2} \lambda_\Phi \Phi \nu^2
\eeq
Then, one finds renormalization group equations (RGEs)
\bea \label{mphiRGE}
\frac{d m_\phi^2}{d \ln Q} &=& \frac{1}{8 \pi^2} |\lambda_\Phi|^2 \left( m_\phi^2 + m_\nu^2 + |A_\Phi|^2 \right)
\\ \label{mnuRGE}
\frac{d m_\nu^2}{d \ln Q} &=& \frac{1}{8 \pi^2} \left[ |\lambda_\Phi|^2 \left( m_\phi^2 + 2 m_\nu^2 + |A_\Phi|^2 \right) + 2 |\lambda_\nu|^2 \left( m_\nu^2 + m_L^2 + m_{H_u}^2 + |A_\nu|^2 \right) \right]
\nonumber \\
&&
\\ \label{aphiRGE}
\frac{d a_\Phi}{d \ln Q} &=& \frac{1}{16 \pi^2} \left( \frac{15}{2} a_\Phi |\lambda_\Phi|^2 + 4 a_\nu \lambda_\Phi \lambda_\nu^* + 2 a_\Phi |\lambda_\nu|^2 \right)
\\ \label{lambdaphiRGE}
\frac{d \lambda_\Phi}{d \ln Q} &=& \frac{1}{16 \pi^2} \lambda_\Phi \left( \frac{5}{2} |\lambda_\Phi|^2 + 2 |\lambda_\nu|^2 \right)
\eea
where $Q$ is the renormalization scale, $m_i^2$ is the soft mass-squared of a scalar field $i$, and $a_i \equiv A_i \lambda_i$ with $A_i$ the $A$-parameter of the trilinear coupling associated with $\lambda_i$.
From \eq{lambdaphiRGE}, \eq{aphiRGE} can be rewritten in terms of $|A_\phi|$ as
\beq \label{AphiRGE}
\frac{d |A_\Phi|^2}{d \ln Q} = \frac{1}{8 \pi^2} \left[ 5 |A_\Phi|^2 |\lambda_\Phi|^2 + 2 \left( A_\Phi A_\nu^* + A_\Phi^* A_\nu \right) |\lambda_\nu|^2 \right]
\eeq

If $\lambda_\Phi \gg \lambda_\nu$, which should be the case for consistency of theory as described in the text, we can ignore all the contributions with $\lambda_\nu$.
Then, Eqs. (\ref{mphiRGE}), (\ref{mnuRGE}), (\ref{AphiRGE}) and (\ref{lambdaphiRGE}) can be approximated to
\bea
\frac{d m_\phi^2}{d \ln Q} &=& \frac{1}{8 \pi^2} |\lambda_\Phi|^2 \left( m_\phi^2 + m_\nu^2 + |A_\Phi|^2 \right)
\\
\frac{d m_\nu^2}{d \ln Q} &=& \frac{1}{8 \pi^2} |\lambda_\Phi|^2 \left( m_\phi^2 + 2 m_\nu^2 + |A_\Phi|^2 \right)
\\
\frac{d |A_\Phi|^2}{d \ln Q} &=& \frac{5}{8 \pi^2} |\lambda_\Phi|^2 |A_\Phi|^2
\\
\frac{d \ln \lambda_\Phi}{d \ln Q} &=& \frac{5}{32 \pi^2} |\lambda_\Phi|^2
\eea
and in order for only $m_\phi^2$ to be negative at low renormalization scale we need 
\beq \label{condformnusq}
\begin{array}{ccc}
m_\phi^2 \ll m_\nu^2 + |A_\phi|^2 &,& m_\nu^2 \gtrsim |A_\Phi|^2
\end{array}
\eeq
For example, for a potential only with the soft-mass term of $\phi$, a numerical analysis shows that input values 
\beq
\begin{array}{ccccc}
m_\nu(M_\planck) = 250 \GeV &,& |A_\Phi(M_\planck)| = 150 \GeV &,& m_\phi(M_\planck) = 100 \GeV
\end{array}
\eeq
with $|\lambda_\Phi(M_\planck)| = 1$ generates vacuum at $\phi_0 \simeq 6 \times 10^{12} \GeV$ where 
\beq
\begin{array}{ccccc}
m_\nu \simeq 210 \GeV &,& |A_\Phi| \simeq 110 \GeV &,& \sqrt{-m_\phi^2} = 12 \GeV
\end{array}
\eeq
with $|\lambda_\Phi| \simeq 0.84$ and $m_\mathrm{PQ} \simeq 23 \GeV$ the physical mass of $\phi$.

\end{document}